\documentclass[%
	conference,
]{IEEEtran}

\usepackage[utf8]{inputenc}
\usepackage{xspace}

\usepackage{standalone}		%
\usepackage{xstring}

\usepackage[binary-units=true]{siunitx}
\usepackage{graphicx}

\usepackage{tikz}
\tikzset{font=\sffamily\footnotesize}

\usetikzlibrary{spy,calc,shapes.arrows,patterns}
\usepackage{pgfplots}
\pgfplotsset{compat=newest}
\pgfplotsset{
    box plot/.style={
        /pgfplots/.cd,
        fill=blue!10,
        only marks,
        mark=-,
        mark size=0.2em,
        /pgfplots/error bars/.cd,
        y dir=plus,
        y explicit,
    },
    box plot box/.style={
        /pgfplots/error bars/draw error bar/.code 2 args={%
            \draw  ##1 -- ++(.2em,0pt) |- ##2 -- ++(-.2em,0pt) |- ##1 -- cycle;
        },
        /pgfplots/table/.cd,
        y index=2,
        y error expr={\thisrowno{3}-\thisrowno{2}},
        /pgfplots/box plot
    },
    box plot top whisker/.style={
        /pgfplots/error bars/draw error bar/.code 2 args={%
            \pgfkeysgetvalue{/pgfplots/error bars/error mark}%
            {\pgfplotserrorbarsmark}%
            \pgfkeysgetvalue{/pgfplots/error bars/error mark options}%
            {\pgfplotserrorbarsmarkopts}%
            \path ##1 -- ##2;
        },
        /pgfplots/table/.cd,
        y index=4,
        y error expr={\thisrowno{2}-\thisrowno{4}},
        /pgfplots/box plot
    },
    box plot bottom whisker/.style={
        /pgfplots/error bars/draw error bar/.code 2 args={%
            \pgfkeysgetvalue{/pgfplots/error bars/error mark}%
            {\pgfplotserrorbarsmark}%
            \pgfkeysgetvalue{/pgfplots/error bars/error mark options}%
            {\pgfplotserrorbarsmarkopts}%
            \path ##1 -- ##2;
        },
        /pgfplots/table/.cd,
        y index=5,
        y error expr={\thisrowno{3}-\thisrowno{5}},
        /pgfplots/box plot
    },
    box plot median/.style={
        /pgfplots/box plot
    },
    boxplot/every median/.style={
    	ultra thick,dashed,cyan
    }
}

\usepackage{todonotes}
\ifCLASSINFOpdf
\else
\fi
\usepackage{amsmath}

\usepackage{sansmath}	%

\usepackage{placeins}

\usepackage[%
IEEE,%
]{copyright-overlay}
\CopyrightAuthor{Robert Falkenberg}
\CopyrightYear{2017}
\ConferenceName{2017 IEEE 85th Vehicular Technology Conference (VTC Spring)}
\Doi{10.1109/VTCSpring.2017.8108515}
\Bibtex{%
@InProceedings\{Falkenberg/etal/2017a,\textCR\textLF
\textHT{}author         = \{Robert Falkenberg and Benjamin Sliwa and Christian Wietfeld\},\textCR\textLF
\textHT{}title          = \{Rushing Full Speed with \{LTE-Advanced\} is Economical - A Power Consumption Analysis\},\textCR\textLF
\textHT{}booktitle      = \{2017 IEEE 85th Vehicular Technology Conference (VTC Spring)\},\textCR\textLF
\textHT{}year           = \{2017\},\textCR\textLF
\textHT{}address        = \{Sydney, Australia\},\textCR\textLF
\textHT{}month          = \{Jun\},\textCR\textLF
\textHT{}doi            = \{10.1109/VTCSpring.2017.8108515\}\textCR\textLF
\}
}

\usepackage[]{acronym}
\usepackage{xspace}

\acrodef{LTE}[LTE]{Long Term Evolution}
\newcommand{\LTE}{\ac{LTE}\xspace}

\acrodef{LTE-A}[LTE-A]{LTE-Advanced}
\newcommand{\LTEA}{\ac{LTE-A}\xspace}

\acrodef{RSSI}[\ensuremath{\mathrm{RSSI}}]{Received Signal Strength Indicator}
\acrodef{RSRP}[\ensuremath{\mathrm{RSRP}}]{Reference Signal Received Power}
\acrodef{RSRQ}[\ensuremath{\mathrm{RSRQ}}]{Reference Signal Received Quality}
\acrodef{SNR}[\ensuremath{\mathrm{SNR}}]{Signal to Noise Ratio}
\acrodef{SINR}[\ensuremath{\mathrm{SINR}}]{Signal to Interference and Noise Ratio}
\acrodef{PRB}[\text{PRB}]{Physical Resource Block}
\acrodef{RB}[\text{RB}]{Resource Block}
\acrodef{CRC}[CRC]{Cyclic Redundancy Check}
\acrodef{DCI}[DCI]{Downlink Control Information}
\acrodef{RNTI}[RNTI]{Radio Network Temporary Identifier}
\acrodef{SI-RNTI}[\ensuremath{\mathrm{SI-RNTI}}]{System-Information \acs{RNTI}}
\acrodef{CCE}[CCE]{Control Channel Element}
\acrodef{PDCCH}[PDCCH]{Physical Downlink Control Channel}
\acrodef{UE}[UE]{User Equipment}
\acrodef{eNodeB}[eNodeB]{evolved NodeB}
\acrodef{SDR}[SDR]{Software Defined Radio}
\acrodef{STG}[STG]{Smart Traffic Generator}
\acrodef{DUT}[DUT]{Device Under Test}
\acrodef{OAI}[OAI]{Open Air Interface}
\acrodef{MCS}[MCS]{Modulation and Coding Scheme}
\acrodef{OFDM}[OFDM]{Orthogonal Frequency Division Multiplexing}
\acrodef{DL}[DL]{Downlink}
\acrodef{UL}[UL]{Uplink}
\acrodef{C3ACE}[C$^3$ACE]{Client-based Control Channel Analysis for Connectivity Estimation}
\acrodef{TCP}[TCP]{Transmission Control Protocol}

\acrodef{FTP}[FTP]{File Transfer Protocol}
\newcommand{\FTP}{\ac{FTP}\xspace}
\acrodef{CoPoMo}{Context-Aware Power Consumption Model}
\newcommand{\copomo}{\ac{CoPoMo}\xspace}
\acrodef{BSE}{Base Station Emulator}
\newcommand{\BSE}{\ac{BSE}\xspace}
\acrodef{CA}{Carrier Aggregation}
\newcommand{\CA}{\ac{CA}\xspace}
\acrodef{CC}{Component Carrier}
\newcommand{\CC}{\ac{CC}\xspace}
\acrodef{MIMO}{Multiple Input Multiple Output}
\newcommand{\MIMO}{\ac{MIMO}\xspace}
\acrodef{SISO}{Single Input Single Output}
\newcommand{\SISO}{\ac{SISO}\xspace}
\acrodef{AWGN}{Additional White Gaussian Noise}
\newcommand{\AWGN}{\ac{AWGN}\xspace}

\acrodef{PCC}[PCC]{Primary Carrier Component}
\newcommand{\PCC}{\ac{PCC}\xspace}
\acrodef{SCC}[SCC]{Secondary Carrier Component}
\newcommand{\SCC}{\ac{SCC}\xspace}

\newcommand{\SNR}{\ac{SNR}\xspace}

\newcommand{\RB}{\ac{RB}\xspace}
\newcommand{\RBs}{\acp{RB}\xspace}

\newcommand{\UE}{\ac{UE}\xspace}
\newcommand{\UEs}{\acp{UE}\xspace}

\newcommand{\DUT}{\ac{DUT}\xspace}

\newcommand{\MCS}{\ac{MCS}\xspace}

\newcommand{\MHz}{\mega\hertz}

\newcommand{\dBm}{dBm}

\newcommand{\DUTA}{DUT-A\xspace}
\newcommand{\DUTB}{DUT-B\xspace}

\newcommand{\Idle}{\textsf{\small{}Idle}\xspace}
\newcommand{\Low}{\textsf{\small{}Low}\xspace}
\newcommand{\High}{\textsf{\small{}High}\xspace}
\newcommand{\Max}{\textsf{\small{}Max}\xspace}

\begin{document}
\title{Rushing Full Speed with LTE-Advanced is Economical - A Power Consumption Analysis}

\author{\IEEEauthorblockN{Robert Falkenberg, Benjamin Sliwa and Christian Wietfeld}
\IEEEauthorblockA{Communication Networks Institute\\TU Dortmund University\\44227 Dortmund, Germany\\
Email: \{Robert.Falkenberg, Benjamin.Sliwa, Christian.Wietfeld\}@tu-dortmund.de}%
}

\maketitle

\begin{abstract}
Boosting data rates in LTE mobile networks is one of the key features of LTE-Advanced.
This improved user experience is achieved by Carrier Aggregation (CA), in which the available spectrum of an operator is bundled out of several frequency bands.
Accordingly, the user equipment has to supply multiple reception chains and therefore consumes considerably more power during a transmission.
On the other hand, transmissions terminate faster, which enables a quick switchover into energy-saving mode.
In order to examine these opposed facts, empirical analyses of existing devices are first carried out.
Subsequently, we present a new CA enhancement of an existing context-aware power consumption model which incorporates the development density of the environment and the mobile device mobility.
Based on the extended model we perform a detailed power consumption analysis and show
that CA leads to power savings of \SI{31}{\percent} if the data rate doubled for large file transmissions.
In addition, we show that CA can lead to power savings even from a data rate increase of \SI{25}{\percent}, regardless of mobility and urban development density.
Besides, the measurement results show that CA operated in the same band leads to a lower power consumption than inter-band CA.

\end{abstract}

\IEEEpeerreviewmaketitle

\section{Introduction}
\begin{figure}[b!]  	
	\centering		  
    \begingroup%
    \StrCount{fig/ca-overview}{/}[\matches]%
    \StrBefore[\matches]{fig/ca-overview}{/}[\datapath]%
    \pgfdeclaredecoration{discontinuity}{start}{
	\state{start}[width=0.5\pgfdecoratedinputsegmentremainingdistance-0.5\pgfdecorationsegmentlength,next state=first wave]
	{}
	\state{first wave}[width=\pgfdecorationsegmentlength, next state=second wave]
	{
		\pgfpathlineto{\pgfpointorigin}
		\pgfpathmoveto{\pgfqpoint{0pt}{\pgfdecorationsegmentamplitude}}
		\pgfpathcurveto
		{\pgfpoint{-0.25*\pgfmetadecorationsegmentlength}{0.75\pgfdecorationsegmentamplitude}}
		{\pgfpoint{-0.25*\pgfmetadecorationsegmentlength}{0.25\pgfdecorationsegmentamplitude}}
		{\pgfpoint{0pt}{0pt}}
		\pgfpathcurveto
		{\pgfpoint{0.25*\pgfmetadecorationsegmentlength}{-0.25\pgfdecorationsegmentamplitude}}
		{\pgfpoint{0.25*\pgfmetadecorationsegmentlength}{-0.75\pgfdecorationsegmentamplitude}}
		{\pgfpoint{0pt}{-\pgfdecorationsegmentamplitude}}
	}
	\state{second wave}[width=0pt, next state=do nothing]
	{
		\pgfpathmoveto{\pgfqpoint{0pt}{\pgfdecorationsegmentamplitude}}
		\pgfpathcurveto
		{\pgfpoint{-0.25*\pgfmetadecorationsegmentlength}{0.75\pgfdecorationsegmentamplitude}}
		{\pgfpoint{-0.25*\pgfmetadecorationsegmentlength}{0.25\pgfdecorationsegmentamplitude}}
		{\pgfpoint{0pt}{0pt}}
		\pgfpathcurveto
		{\pgfpoint{0.25*\pgfmetadecorationsegmentlength}{-0.25\pgfdecorationsegmentamplitude}}
		{\pgfpoint{0.25*\pgfmetadecorationsegmentlength}{-0.75\pgfdecorationsegmentamplitude}}
		{\pgfpoint{0pt}{-\pgfdecorationsegmentamplitude}}
		\pgfpathmoveto{\pgfpointorigin}
	}
	\state{do nothing}[width=\pgfdecorationsegmentlength,next state=do nothing]{
		\pgfpathlineto{\pgfpointdecoratedinputsegmentlast}
	}
	\state{final}
	{
		\pgfpathlineto{\pgfpointdecoratedpathlast}
	}
}

\begin{tikzpicture}[font=\sffamily\footnotesize]
\begin{sansmath}

\def\sxa{2cm}
\def\sya{3cm}
\def\sxb{6cm}
\def\syb{3cm}
\tikzstyle{axraw} = [thick]
\tikzstyle{ax} = [axraw, -stealth]
\tikzstyle{spectrum} = [draw, minimum height=1cm, minimum width=1cm, inner sep=0, anchor=south]
\tikzstyle{label} = [node distance=0.5cm]

\newcommand{\speedmeter}[3]{
	\def\meterradius{0.7cm}
	\def\meterunit{#1}
	\def\metervalue{#2}
	\def\meterredstart{30}
	\def\meterredend{-45}
	\coordinate (metercoordinate) at #3;
	\draw[] (metercoordinate) circle (\meterradius);
	\foreach \x in {215, 200, ..., -45} {\draw[] ([shift={(metercoordinate)}]\x:\meterradius-0.15cm) -- ([shift={(metercoordinate)}]\x:\meterradius-0.05cm);}
	\draw[fill=black] ([shift={(metercoordinate)}]215:\meterradius-0.2cm) circle (0.5pt);
	\draw[red, thick] ([shift={(metercoordinate)}]\meterredstart:\meterradius-0.05cm) arc (\meterredstart:\meterredend:\meterradius-0.05cm);
	\draw[-,very thick] ([shift={(metercoordinate)}]\metervalue:\meterradius-0.05cm) to[] ([shift={(metercoordinate)}]\metervalue:-1mm);
	\draw[fill=white] (metercoordinate) circle (1pt);
	\node[inner sep=0pt, below=2mm of metercoordinate] {\footnotesize\meterunit};
}

\coordinate (a1) at (0cm,0cm);
\coordinate (a2) at (4cm,0cm);

\draw[ax] (a1)  to node[at end, below=1mm,anchor=north east, inner sep=0pt] {Frequency}  +(3cm, 0cm);
\draw[ax] (a1)  to node[midway, above, rotate=90] {Signal} +(0cm, 1.1cm);
\node[spectrum, fill=blue!30, draw=blue] (pcc1)  at ([xshift=1cm]a1) {};

\path[] (a2)  to node[at end, below=1mm,anchor=north east, inner sep=0pt] {Frequency}  +(3cm, 0cm);
\draw[axraw] (a2) to +(1.3cm, 0cm);
\draw[ax] ([shift={(1.7cm, 0cm)}]a2) to +(1.3cm, 0cm);
\draw[thick, decoration={%
	discontinuity,
	amplitude=0.1cm,
	segment length=0.1cm,
	meta-segment length=0.2cm},
decorate] ([shift={(1.3cm, 0cm)}]a2) -- ([shift={(1.7cm, 0cm)}]a2);

\draw[ax] (a2)  to node[midway, above, rotate=90] {Signal} +(0cm, 1.1cm);
\node[spectrum, fill=blue!30, draw=blue] (pcc2) at ([xshift=0.8cm]a2) {};
\node[spectrum, fill=red!30, draw=red] (scc2) at ([xshift=2.2cm]a2) {};

\node[label, below=of pcc1, text width=2.5cm, align=center] (pcc1l) {Single Carrier Reception};
\draw[-stealth] (pcc1l) -- (pcc1);

\node[label, below=of pcc2, text width=2.6cm, align=center] (pcc2l) {Multiple Carrier Reception};
\draw[-stealth] (pcc2l) -- (pcc2);
\draw[-stealth] (pcc2l) -- (scc2);

\speedmeter{P}{120}{(0.5cm,2.6cm)}
\speedmeter{Mbps}{120}{(2.0cm,2.6cm)}
\node[] at (0.5, 1.7) {Power};
\node[] at (2.0, 1.7) {Speed};

\speedmeter{P}{90}{(4.5cm,2.6cm)}
\speedmeter{Mbps}{10}{(6.0cm,2.6cm)}
\node[] at (4.5, 1.7) {Power};
\node[] at (6.0, 1.7) {Speed};

\node[] at (1.5cm,4cm) {\textbf{LTE}};
\node[text width=4cm, align=center] at (5.5cm, 4cm) {\textbf{LTE-Advanced\\Carrier Aggregation}};

\end{sansmath}
\end{tikzpicture}%
    \endgroup%

	\vspace{-15pt}
	\caption{Functional Principle of Carrier Aggregation and its Influence on Data Rate and Power Consumption}
	\label{fig:overview}
	\vspace{-10pt}
\end{figure}
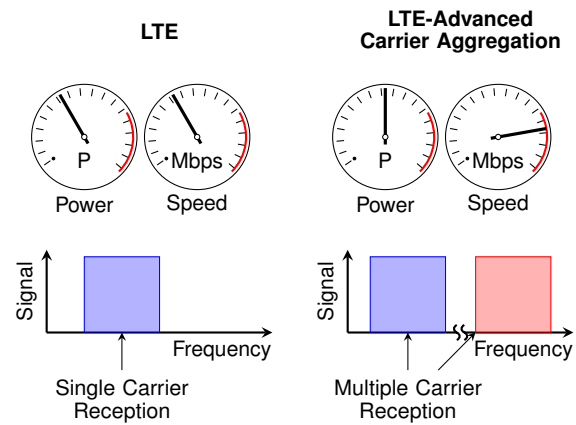

\PrintCopyrightOverlay
The ongoing evolution of \LTE mobile networks increases the achievable data rate over the radio link.
A prominent example is the \acf{CA} feature of \LTEA networks in which, 
as shown in \figurename{~\ref{fig:overview}},
the data is transmitted simultaneously over multiple distributed frequency bands possessed by a single network operator in order to boost the data rate of the connection.
While data rates increased in the recent years, in spite of more and more efficient system-on-chips and new power-saving techniques, the battery lifetimes of mobile devices did not.
This is caused by the growing demands on the provided content: 
the resolution of video streams increases, websites and documents contain more complex graphical components, and the amount of data which is being stored and processed by cloud services grows continually.
As the lion share of today's smartphone traffic is issued in the downlink due to streaming and downloading web content, operators in Europe began to roll out \LTEA networks supporting \CA in the downlink.

For an \LTEA modem the reception of multiple carriers requires powering a corresponding number of receive chains and additional signal processing.
At this point arises the question how the power consumption of an \LTEA device is affected by \CA.
On the one hand, as illustrated in \figurename{~\ref{fig:overview}}, \CA leads to an increased power draw within the \UE.
On the other hand, the data rate is significantly boosted, hence allowing a faster termination of transmissions.

The objective of this work is to analyze the power consumption of today's \LTEA devices and to utilize this empirical knowledge for consumption simulations in numerous usage patterns, environments, and mobility scenarios, e.g. for a moving vehicle.
This allows the identification of situations where \CA reduces the power draw of \UE and therefore increases the battery lifetime of mobile devices.

For this purpose, 
an established power consumption model will be extended for downlink \CA in Sec.~\ref{ch:model}.
Afterwards, Sec.~\ref{ch:measurements} covers empirical measurements with existing \LTEA devices followed by various simulations in Sec.~\ref{ch:simulations} which are based on the gained values and the proposed model extension.
The paper closes with a brief summary and conclusion on the results.

\section{Related Work}\label{ch:related}
The power consumption of \LTE equipment has been measured, modeled and analyzed in numerous studies.
In \cite{Jensen2012} a power consumption model for downlink and uplink is presented based on empirical measurements.
It shows that the power consumption for the uplink is mostly affected by the transmission power rather than by the uplink data rate.
Complementary, the power draw invoked by downlink reception mostly depends on the data rate rather than received signal power.
The most complex procedure in the reception process is the iterative turbo decoding whose complexity linearly scales with the data rate.
This fact was confirmed in \cite{Lauridsen2014} as the downlink power consumption induced by the turbo decoder of newer devices linearly grows
by \SI{5}{\percent} if the number of allocated \RBs increases by factor \num{10}.
In case of an \LTE cell with multiple \UEs claiming for a large number of radio resources (e.g., large file transfers) it is energy efficient to share the available \RBs in time rather than in frequency~\cite{PhdDusza2013}.
By doing so, single \UEs process larger chunks of data for a short time and save energy in the remaining period instead of powering the decoder continuously.

Moreover, the authors of \cite{Lauridsen2014} attend on the energy-efficiency evolution in the past few years.
They point out a reduction of transmission-power consumption by \SI{35}{\percent} at low and intermediate output power levels as well as for the entire reception process.
Besides, the uplink power consumption of \LTE \UE  is independent from the actual uplink data rate.

In \cite{Sanchez2014} an empirical analysis of the current consumption of first-generation \acl{LTE-A} devices has been performed.
The authors have measured power savings of \SI{13}{\percent} by enabling \CA for large \FTP downloads.
The authors of \cite{Sonkusare2015} were able to confirm the same effect by their measurements.
Due to restrictions of the \DUT in both studies, no intra-band \CA could be analyzed,
so our measurements will close this gap.

Since laboratory measurements typically lack of generality and field measurements often cannot be performed in the required extent, a context aware power consumption model has been developed in~\cite{DuszaJournal2013} to perform representative simulations of the \UE's power consumption.
The model considers the radio channel (static, fading), environment (urban, rural), data size, arrival rate, \UE power consumption in different operation modes, as well as further parameters.
From these parameters the model calculates the average long-term power consumption of the \UE.
The development and evaluation of the model showed that the transmission of data in uplink direction is the most energy-intensive part of an \LTE modem -- especially at high transmission power.

Furthermore, the model has been extended for \LTEA uplink analysis \cite{Dusza2014} and analyzed in theory.
To the best of our knowledge the model has not been adapted for an \LTEA downlink analysis, which will be proposed in this paper.

\section{Power Consumption Model}\label{ch:model}
Performing representative simulations on the impact of \CA on the \UE's power consumption requires an exact power model of the analyzed device as well as an incorporation of external influencing factors like the environment and mobility.
For this reason the proposed power consumption model for downlink \CA is based on the excessively studied \copomo presented in~\cite{DuszaJournal2013}.
In order to facilitate a better understanding of the presented approach, an introduction to the underlying model is given in the next section.
For full details the reader is kindly referred to the full documentation in~\cite{PhdDusza2013} and \cite{DuszaJournal2013}.

\subsection{Uplink Power Consumption Model}\label{ch:copomo}
One central statement of \copomo is the outcome
that the amount of power an \LTE modem requires is highly depending on its transmission power and the time spent in this mode.
The authors have reduced the complexity of uplink power consumption to four main power states (\Idle, \Low, \High, \Max) a \UE may enter according to the current radio conditions while still maintaining a very high model accuracy.
Each state $i$ is characterized by an average power consumption $\bar{P}_i$ obtained from empirical measurements of the actually analyzed device (cf.~\ref{ch:measurements}).
Furthermore, downlink transmissions without any uplink component result in entering the \Low power state regardless of the radio conditions.
Those states, in conjunction with corresponding transition rates, build the midpoint of the power model and can be expressed as a Markov chain as shown in \figurename{\ref{fig:statemachine}}.
Modeling the idle time between two uplink transmissions as a negative exponential distributed function with mean $1/\lambda$ leads to the leaving transitions from \Idle state with the rate $\lambda$.
The particular rates $\lambda_i$ are the result of weighting $\lambda$ with individual power state probabilities $\vartheta_i$:
\begin{eqnarray}
\lambda_i &=& \vartheta_i \cdot \lambda\qquad\text{for } i \in \lbrace 2,3,4 \rbrace,\\
\sum_{i=2}^{4} \vartheta_i &=& 1.
\end{eqnarray}
They reflect the probability that the device must transmit at a specific transmission power to achieve the required \SNR at the addressed base station under the given radio conditions.
The actual values of $\vartheta_i$ depends on the cell environment, e.g., urban or rural area, the \UE's mobility, e.g., pedestrian or vehicular movement, and finally the utilized carrier frequency.
In this paper, and according to~\cite{DuszaJournal2013},
$\vartheta_i$ is derived from ray tracing simulations and channel emulations of three environments, three mobility patterns, and two common carrier frequencies as listed in Tab.~\ref{tab:raytracing}.

\begin{figure}[tb]  	
	\centering		  
    \begingroup%
    \StrCount{fig/statemachine}{/}[\matches]%
    \StrBefore[\matches]{fig/statemachine}{/}[\datapath]%
    \begin{tikzpicture}[]
\begin{sansmath}

\colorlet{TxLow}{yellow!30}
\colorlet{TxHigh}{orange!60}
\colorlet{TxMax}{red!80}
\colorlet{Rx}{green!30}
\colorlet{Idle}{white}

\tikzstyle{state} = [draw=black, circle, inner sep=0pt, outer sep=0pt, node distance=0.5cm, minimum height=1cm, text width=1cm, align=center]
\tikzstyle{idle} = [state]
\tikzstyle{TxL} = [state, fill=TxLow]
\tikzstyle{TxH} = [state, fill=TxHigh]
\tikzstyle{TxM} = [state, fill=TxMax]
\tikzstyle{edge} = [-stealth, bend left=45]
\tikzstyle{label} = [above, at end]
\tikzstyle{ulabel} = [below, at start]

\node[idle] (i) {1\\Idle};
\node[TxL, right=of i] (l) {2\\Low};
\node[TxH, right=of l] (h) {3\\High};
\node[TxM, right=of h] (m) {4\\Max};

\draw[edge] (i) to node[label] {$\lambda_2$} (l);
\draw[edge] (i) to node[label] {$\lambda_3$} (h);
\draw[edge] (i) to node[label] {$\lambda_4$} (m);

\draw[edge] (l) to node[ulabel] {$\mu_2$} (i);
\draw[edge] (h) to node[ulabel] {$\mu_3$} (i);
\draw[edge] (m) to node[ulabel] {$\mu_4$} (i);

\end{sansmath}
\end{tikzpicture}%
    \endgroup%

	\vspace{-10pt}
	\caption{Markovian Power State Model of LTE User Equipment}
	\label{fig:statemachine}
	\vspace{-5pt}
\end{figure}
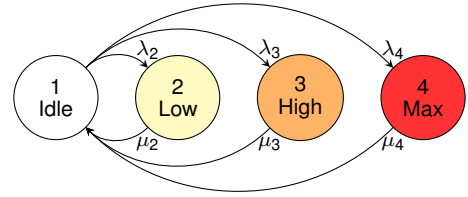
\begin{table}[!t]
	\renewcommand{\arraystretch}{1.3}
	\caption{Properties of the Simulated Environments According~to~\cite{DuszaJournal2013}~if~not~Noted~Otherwise}
	\label{tab:raytracing}
	\centering
	\begin{tabular}{lr}
		\hline
		Environment & Inter Site Distance\\
		\hline
		Urban & \SI{500}{\meter} \cite{25.813} \\
		Suburban & \SI{1732}{\meter} \cite{25.813} \\
		Rural & \SI{5000}{\meter}\\
		\hline
		Channel model & Mobility\\
		\hline
		\AWGN & \SI{0}{\kilo\meter\per\hour}\\
		Pedestrian & \SI{3}{\kilo\meter\per\hour}\\
		Vehicular & \SI{60}{\kilo\meter\per\hour}\\
		\hline
		Frequency Band & Bandwidth \\
		\hline
		\SI{800}{\MHz} (Band 20) & \SI{10}{\MHz}\\
		\SI{2600}{\MHz} (Band 7) & \SI{10}{\MHz}\\
		\hline
	\end{tabular}
	\vspace{-5pt}
\end{table}

The time spent in any active state $1/\mu_{\text{UL},i}$ is derived from the channel dependent uplink throughput in each state $R_{\text{UL}, i}$ and the average upload file size $D_{\text{UL}}$:
\begin{eqnarray}
\mu_{\text{UL},i} &=& \frac{R_i}{D_\text{UL}} \qquad\text{for } i \in \lbrace 2,3,4 \rbrace.
\end{eqnarray}
Especially in the \Max state, where the maximum transmission power is limited to \SI{23}{\dBm}~\cite{36.101} and the \UE must fall back to more robust \MCS, $R_{\text{UL}, 4}$ may be significantly lower than in the other modes.
The model considers this rate degradation according to empirical studies performed in~\cite{Dusza2012_5} and~\cite{Dusza2013a}, which leads to a longer residence time in \Max state.

By deriving the equilibrium equations for the Markov model in \figurename{~\ref{fig:statemachine}} and solving the linear equation system we get the state probabilities $p_i$ with
\begin{eqnarray}
p_i = \begin{cases}
\frac{1}{1 + \sum\limits_{j=2}^{4} (\lambda_j/\mu_j)} & \text{for } i=1, \\
\frac{\lambda_i/\mu_i}{1 + \sum\limits_{j=2}^{4} (\lambda_j/\mu_j)} & \text{for } i\not =1.\label{eq:equilibrium}
\end{cases}
\end{eqnarray}
With this state probabilities and the average power consumption in each state $\bar{P}_i$ the average long term power consumption of the \UE can be expressed as
\begin{eqnarray}
P_{\text{Total}} &=& \sum_{i=1}^{4} \bar{P}_i \cdot p_i.
\end{eqnarray}

\subsection{Model Extension for Downlink and Carrier Aggregation}\label{ch:camodel}
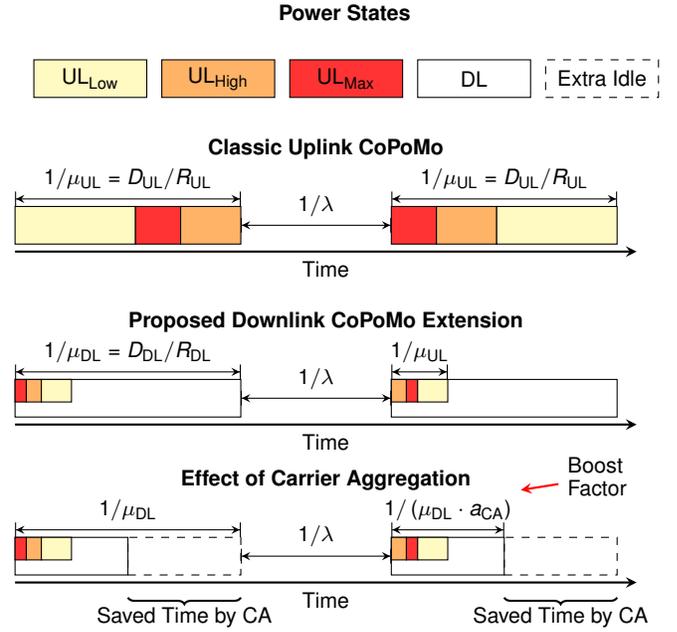
\begin{figure}[tb]  	
	\centering		  
    \begingroup%
    \StrCount{fig/timeline}{/}[\matches]%
    \StrBefore[\matches]{fig/timeline}{/}[\datapath]%
    \begin{tikzpicture}[font=\sffamily\footnotesize]
\begin{sansmath}

\def\axshift{-1mm}
\def\slot{1mm}
\def\vgap{1.8cm}		%
\def\vpadding{1mm}	%
\def\vpaddingB{3mm}	%
\def\captiondist{1.1cm}	%
\colorlet{TxLow}{yellow!30}
\colorlet{TxHigh}{orange!60}
\colorlet{TxMax}{red!80}
\colorlet{Rx}{white}
\colorlet{Idle}{white}

\tikzstyle{timeline} = [draw, thick, -stealth]
\tikzstyle{interval} = [draw, |{stealth}-{stealth}|]
\tikzstyle{caption} = [font=\bfseries\sffamily\footnotesize]
\tikzstyle{box} = [draw=black, inner sep=0pt, outer sep=0pt, node distance=0pt, minimum height=0.5cm]
\tikzstyle{TxL} = [box, fill=TxLow]
\tikzstyle{TxH} = [box, fill=TxHigh]
\tikzstyle{TxM} = [box, fill=TxMax]
\tikzstyle{RxTxL} = [TxL, minimum height=0.3cm]
\tikzstyle{RxTxH} = [TxH, minimum height=0.3cm]
\tikzstyle{RxTxM} = [TxM, minimum height=0.3cm]
\tikzstyle{Rx} = [box, fill=Rx]
\tikzstyle{Idle} = [box, dashed]
\tikzstyle{legend} = [node distance=2mm, minimum width=1.5cm]
\node[TxL, legend] (legL)  at(1cm,2.2cm){$\text{UL}_{\text{Low}}$};
\node[TxH, legend, right=of legL] (legH){$\text{UL}_{\text{High}}$};
\node[TxM, legend, right=of legH] (legM){$\text{UL}_{\text{Max}}$};
\node[Rx, legend, right=of legM](legR){DL};
\node[Idle, legend, right=of legR](legI){Extra Idle};
\path[] ([yshift=\axshift]legL.south west) to node[caption, above=1cm] (ps) {Power States} +(235.0pt,0pt);
\node[text=white, below=0.0cm of ps, font=\tiny\itshape] {For my lovely daughter Vivien Falkenberg};

\node[TxL, anchor=south west, minimum width=16*\slot]  (UL1) at (0cm,0cm)  {};
\node[TxM, minimum width=6*\slot, right=of UL1]  (UL2) {};
\node[TxH, minimum width=8*\slot, right=of UL2]  (UL3) {};
\node[TxM, minimum width=6*\slot, right=20*\slot of UL3] (UL4) {};
\node[TxH, minimum width=8*\slot, right=of UL4] (UL5) {};
\node[TxL, minimum width=16*\slot, right=of UL5] (UL6) {};
\draw[timeline] ([yshift=\axshift]UL1.south west) to node[caption, above=\captiondist] {Classic Uplink CoPoMo} node[below, midway] {Time} +(235.0pt,0pt);
\draw[interval] (UL3.east) to node[midway, above] {$1/\lambda$} (UL4.west);
\draw[interval] ([yshift=\vpadding]UL1.north west) to node[midway, above] {$1/\mu_{\text{UL}} = D_{\text{UL}}/R_{\text{UL}}$} ([yshift=\vpadding]UL3.north east);
\draw[interval] ([yshift=\vpadding]UL4.north west) to node[midway, above] {$1/\mu_{\text{UL}} = D_{\text{UL}}/R_{\text{UL}}$} ([yshift=\vpadding]UL6.north east);

\node[Rx, below=\vgap of UL1.south west, anchor=north west, minimum width=30*\slot](RX1){};
\node[RxTxM, anchor=north west, minimum width=1.5*\slot] at (RX1.north west) (UUL1){};
\node[RxTxH, minimum width=2*\slot, right=of UUL1](UUL2){};
\node[RxTxL, minimum width=4*\slot, right=of UUL2](UUL3){};

\node[Rx, minimum width=30*\slot, right=20*\slot of RX1](RX2){};
\node[RxTxH, anchor=north west, minimum width=2*\slot] at (RX2.north west) (UUL4){};
\node[RxTxM, minimum width=1.5*\slot, right=of UUL4](UUL5){};
\node[RxTxL, minimum width=4*\slot, right=of UUL5](UUL6){};

\draw[timeline] ([yshift=\axshift]RX1.south west) to node[caption, above=\captiondist] {Proposed Downlink CoPoMo Extension} node[below, midway] {Time} +(235.0pt,0pt);
\draw[interval] (RX1.east) to node[midway, above] {$1/\lambda$} (RX2.west);
\draw[interval] ([yshift=\vpadding]UUL4.north west) to node[midway, above] {$1/\mu_{\text{UL}}$} ([yshift=\vpadding]UUL6.north east);
\draw[interval] ([yshift=\vpadding]UUL1.north west) to node[midway, above] {$1/\mu_{\text{DL}} = D_{\text{DL}}/R_{\text{DL}}$} ([yshift=\vpadding]RX1.north east);

\node[Rx, below=\vgap of UUL1.south west, anchor=north west, minimum width=15*\slot](CARX1){};
\node[RxTxM, anchor=north west, minimum width=1.5*\slot] at (CARX1.north west) (CAUL1){};
\node[RxTxH, minimum width=2*\slot, right=of CAUL1](CAUL2){};
\node[RxTxL, minimum width=4*\slot, right=of CAUL2](CAUL3){};
\node[Idle, minimum width=15*\slot, right=of CARX1](CAI1){};

\node[Rx, right=20*\slot of CAI1, minimum width=15*\slot](CARX2){};
\node[RxTxH, anchor=north west, minimum width=2*\slot] at (CARX2.north west)(CAUL4){};
\node[RxTxM, minimum width=1.5*\slot, right=of CAUL4](CAUL5){};
\node[RxTxL, minimum width=4*\slot, right=of CAUL5](CAUL6){};
\node[Idle, minimum width=15*\slot, right=of CARX2](CAI2){};
\draw[timeline] ([yshift=\axshift]CARX1.south west) to node[caption, above=\captiondist] {Effect of Carrier Aggregation} node[below, midway] {Time} +(235.0pt,0pt);
\draw[interval] (CAI1.east) to node[midway, above] {$1/\lambda$} (CARX2.west);
\draw[interval] ([yshift=\vpadding]CAUL1.north west) to node[midway, above] {$1/\mu_{\text{DL}}$} ([yshift=\vpadding]CAI1.north east);
\draw[interval] ([yshift=\vpadding]CAUL4.north west) to node[midway, above] (gg) {$1/\left(\mu_{\text{DL}}\cdot a_{\text{CA}}\right)$} ([yshift=\vpadding]CARX2.north east);
\draw[thick, decorate,decoration={brace, mirror}] ([yshift=-\vpaddingB]CAI1.south west) to node[midway, below] {Saved Time by CA} ([yshift=-\vpaddingB]CAI1.south east);
\draw[thick, decorate,decoration={brace, mirror}] ([yshift=-\vpaddingB]CAI2.south west) to node[midway, below] {Saved Time by CA} ([yshift=-\vpaddingB]CAI2.south east);
\node[above right=1mm and 5mm of gg.north east, text width=1cm, yshift=-3mm] (hh) {Boost Factor};
\draw[-stealth, thick, red] (hh) -- (gg.north east);

\end{sansmath}
\end{tikzpicture}%
    \endgroup%

	\vspace{-5pt}
	\caption{Comparison of Uplink CoPoMo and the Proposed Extension for Downlink and Carrier Aggregation}
	\label{fig:timeline}
	\vspace{-5pt}
\end{figure}

In order to apply \copomo to mixed uplink/downlink transmissions and carrier aggregation the model must be extended. The necessary modifications will be explained in this subsection.
A timeline of an exemplary uplink transmission according to classic \copomo and \figurename{~\ref{fig:statemachine}} is shown in the upper part of \figurename{~\ref{fig:timeline}}.
It shows the residence time in each state according to calculated long-term state probabilities from Eq.~\ref{eq:equilibrium}.
Although the model explicitly does not include transitions between two distinct transmitting states (cf. \figurename{~\ref{fig:statemachine}}), those states are still grouped next to each other in the figure.
This is legitimate, because the power consumption of a device is assumed to be independent of the order the states are traversed.
Therefore, only the average time spent in each state must be maintained.

\begin{figure*}[!h!t!]
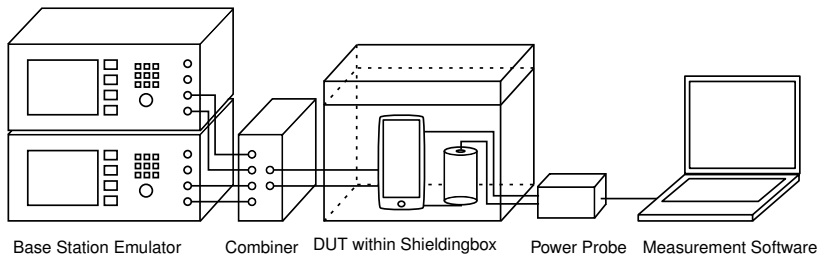
  	
	\centering		  
	\scalebox{0.8}{\includestandalone{fig/meas-setup}}\hspace*{1cm}
	\scalebox{0.8}{%
    \begingroup%
    \StrCount{fig/meas-photo}{/}[\matches]%
    \StrBefore[\matches]{fig/meas-photo}{/}[\datapath]%
    \includestandalone[]{fig/meas-photo}%
    \endgroup%
}
	\vspace{-5pt}
	\caption{Schematic Overview of the Measurement Setup and Photography of the Measurement Setup Inside a Shielded Chamber}
	\label{fig:setup}
	\vspace{-10pt}
\end{figure*}

To extend the power model for downlink transmissions the following assumptions were made:
\begin{itemize}
	\item The data rates of the uplink $R_{\text{UL}}$ and the downlink $R_{\text{DL}}$ are affected by the radio channel in the same way. Therefore, the relation between the two rates $r_{\text{DLUL}}$ is fixed to
			$r_{\text{DLUL}} = R_{\text{UL}}/R_{\text{DL}}$.
	\item Realistic downlink transmissions typically contain a fraction of uplink data, e.g. requests and acknowledgments. The relation of downlink data size $D_{\text{DL}}$ and uplink data size $D_{\text{UL}}$ is described by
			$d_{\text{DLUL}} = D_{\text{UL}}/D_{\text{DL}}$.
	\item The arrangement of uplink transmissions during an ongoing download has no effect on the power consumption. It is assumed, that no additional time is spent in tail states~\cite{PhdDusza2013}.
	\item Downloading data during an active uplink transmission does not increase the average power consumption of the device in the current active state.%
	\item The power consumption during a pure downlink transmission $\bar{P}_\text{DL}$ (without any uplink) equals to the power consumption in \Low mode as shown in previous studies~\cite{PhdDusza2013}. Therefore, 
		$\bar{P}_\text{DL} = \bar{P}_2$.
	\item Carrier aggregation will be modeled as a boost factor $a_\text{CA} \geq 1$ that increases the downlink data rate $R_\text{DL}$. As measurements in Sec.~\ref{ch:measurements} will show, enabling an additional \CC leads to a constantly higher power consumption in all active states.
\end{itemize}
With these assumptions a downlink transmission without \CA ($a_\text{CA}=1$) can be arranged in a time line as shown in the middle of \figurename{~\ref{fig:timeline}}. The example shows two downlink transmissions (white) with an associated uplink portion (yellow, orange and red) arranged to the beginning of each block.
By using $\mu_{\text{DL}}$ instead of $\mu_{\text{UL}}$ to calculate the stationary distribution of the Markov chain (Eq.~\ref{eq:equilibrium}), the resulting state probabilities reflect the relative stay time of each mode as if the device was continuously uploading during the entire downlink transmission period of
\begin{eqnarray}
\frac{1}{\mu_{\text{DL}}} = \frac{D_{\text{DL}}}{R_{\text{DL}}} = \frac{D_{\text{UL}}\cdot r_{\text{DLUL}}}{R_{\text{UL}}\cdot d_{\text{DLUL}}}.
\end{eqnarray}
As the actual uplink time during the particular power states is only $1/\mu_{\text{UL}}$, the remaining time consists of pure downlink traffic consuming $\bar{P}_{\textsc{DL}}$.

When enabling an additional \CC and therefore increasing the downlink data rate by $a_\text{CA} > 1$ this will result in a shorter downlink transmission time as shown in the bottom row of \figurename{~\ref{fig:timeline}}.
During this saved time the \UE immediately enters the \Idle state, thus saves energy.

Let $t_i = 1/\mu_{\text{DL},i}$ the average time spent in a particular state $i$ without \CA and $t_{\text{ULDL},i} = 1/(\mu_{\text{DL},i}\cdot a_\text{CA})$ the actual active time with \CA then the extra idle time gained by \CA is
\begin{eqnarray}
t_{\text{idle},i} &=& t_i - t_{\text{ULDL},i}.
\end{eqnarray}
The average uplink time is derived from its data size and rate
$t_{\text{UL},i} = D_{\text{UL}}/R_{\text{UL}, i}$
and the remaining active time consists of only pure downlink transmissions:
\begin{eqnarray}
t_{\text{DL},i} &=&  t_{\text{ULDL},i} - t_{\text{UL},i} \qquad \text{for } t_{\text{ULDL},i} > t_{\text{UL},i}.
\end{eqnarray}
Cases where the uplink transmission time is larger than or equal to the downlink activity time are covered by the classic \copomo, thus are not further discussed in this paper.

According to the calculated time intervals, the mixed power consumption $\tilde{P}_i$ can be described as
\begin{eqnarray}
\tilde{P}_i = 
\frac{t_{\text{UL}, i}}{t_i} \bar{P}_i +
\frac{t_{\text{DL}, i}}{t_i} \bar{P}_\text{DL} +
\frac{t_{\text{idle}, i}}{t_i} \bar{P}_1  & \text{for } i \in \lbrace 2,3,4\rbrace \label{eq:capower}
\end{eqnarray}
and $\tilde{P}_1 = \bar{P}_1$.
This model now allows a simulative rating of the impact of \CA on the \UE power consumption pursuant to different system and context parameters.

\section{Measurements}\label{ch:measurements}
In order to fed the proposed power consumption model with appropriate values and to analyze the static power consumption caused by downlink \CA in today's \UE,
we performed excessive measurements with two Common-off-the-shelf \LTE \acf{DUT} in a laboratory environment, abbreviated as \DUTA and \DUTB:
\begin{itemize}
	\item \DUTA: LG G5
	\item \DUTB: Samsung Galaxy S5 Neo
\end{itemize}

\subsection{Measurement Setup}
A measurement setup, as outlined in \figurename{~\ref{fig:setup}}, was built to create a fully controlled environment consisting of an \LTE \BSE, an \LTE \UE, and measurement equipment to determine the \UE power consumption.
The \BSE was a composite of two \emph{Rohde \& Schwarz CMW500} that allows the emulation of an \LTE cell with up to three \acp{CC} à \SI{20}{\mega\hertz} at arbitrary frequencies and using $2\times2$ \MIMO on each carrier.
Due to limitations of the available \DUT, measurements with \CA were limited to the \PCC (the actual serving cell) and a single \SCC which formed the aggregated part.
The signal was fed into a shielded chamber as shown in \figurename{~\ref{fig:setup}} to avoid interference with public networks.
In order to measure the power consumption of the \DUT, the battery has been extracted and replaced by wired dummy which was interconnected with a measurement probe.
The power consumption was captured by a \emph{Hitex Powerscale} with a sampling rate of \SI{100}{\kilo\hertz} and evaluated on a computer.

If not stated otherwise the measurements were performed in \SISO configuration at bandwidths of \SI{10}{\mega\hertz} for each \CC.
During an active connection the \UE continuously got full \RB allocations on all active \CC forcing the \UE to decode random padding bits and withdraw them in downlink direction and to fill padding bits in the uplink. %
By doing so, no user-space application was required on the \DUT to generate traffic.
This minimized the influence of other power consuming components (e.g. CPU, memory, display) on measurements of the \LTE part, as only the over-all power consumption of the \DUT could be measured.

\subsection{Band Specific Power Consumption}
\begin{figure}[tb]  	
	\centering		  
    \begingroup%
    \StrCount{fig/additional-CA-power-G5}{/}[\matches]%
    \StrBefore[\matches]{fig/additional-CA-power-G5}{/}[\datapath]%
    \begin{tikzpicture}[spy using outlines={circle, magnification=4, connect spies}, font=\sffamily\footnotesize]
\begin{sansmath}

\pgfplotstableread[row sep=\\,col sep=&]{
	PCC & SCC800 & SCC1800 & SCC2600 \\
	800     & 0  & 273  & 323  \\
	1800     & 300 & 190  & 310  \\
	2600    & 310 & 310 & 190 \\
}\mydataG

\pgfplotstableread[row sep=\\,col sep=&]{
	PCC & SCC800 & SCC1800 & SCC2600 \\
	800     & 300  & 339  & 403  \\
	1800     & 314 & 307  & 394  \\
	2600    & 330 & 310 & 330 \\
}\mydataS

    \begin{axis}[
	ybar,
	width= 235.0pt,			%
	height=5.5cm,
	enlarge x limits=0.3,
	ymin=0, ymax=500,
    symbolic x coords={800,1800,2600},
    bar width=0.5cm,
    xtick=data,
    nodes near coords,
	xlabel={Primary Component Carrier Frequency [\si{\mega\hertz}]},
	ylabel={Power Consumption [\si{\milli\watt}]},
	legend cell align=left,
 	legend pos=south east,
legend style={legend columns=-1,anchor=south east},
    ]
    \addplot[black, fill=white] table[x=PCC,y=SCC800]{\mydataG};
    \addplot[black, fill=gray] table[x=PCC,y=SCC1800]{\mydataG};
    \addplot[black, fill=black] table[x=PCC,y=SCC2600]{\mydataG};
    \legend{SCC 800, SCC 1800, SCC 2600}
    \end{axis}

	\node[draw, black, rotate=90, anchor=west, fill=white, font=\tiny, minimum height=0.5cm] at (0.7cm,0cm) {Not Supported};
	
	\node[draw] (label) at (3.4cm,3.1cm) {Intra-Band CA};
	\draw[-stealth] (label) -| +(-2.7cm,-1.0cm);
	\draw[-stealth] (label) -| +(2.6cm,-1.0cm);
	\draw[-stealth] (label) -- +(0cm,-1.0cm);
	
	\node[draw, fill=white] at (-4mm,-4mm) {DUT-A};

\end{sansmath}
\end{tikzpicture}%
    \endgroup%

	\vspace{15pt}
    \begingroup%
    \StrCount{fig/additional-CA-power-S5}{/}[\matches]%
    \StrBefore[\matches]{fig/additional-CA-power-S5}{/}[\datapath]%
    \begin{tikzpicture}[spy using outlines={circle, magnification=4, connect spies}, font=\sffamily\footnotesize]
\begin{sansmath}

\pgfplotstableread[row sep=\\,col sep=&]{
	PCC & SCC800 & SCC1800 & SCC2600 \\
	800     & 0  & 273  & 323  \\
	1800     & 300 & 190  & 310  \\
	2600    & 310 & 310 & 190 \\
}\mydataG

\pgfplotstableread[row sep=\\,col sep=&]{
	PCC & SCC800 & SCC1800 & SCC2600 \\
	800     & 300  & 339  & 403  \\
	1800     & 314 & 307  & 394  \\
	2600    & 330 & 310 & 330 \\
}\mydataS

    \begin{axis}[
	ybar,
	width= 235.0pt,			%
	height=5.5cm,
	enlarge x limits=0.3,
	ymin=0, ymax=550,
    symbolic x coords={800,1800,2600},
    bar width=0.5cm,
    xtick=data,
    nodes near coords,
	xlabel={Primary Component Carrier Frequency [\si{\mega\hertz}]},
	ylabel={Power Consumption [\si{\milli\watt}]},
	legend cell align=left,
 	legend pos=south east,
 	legend style={legend columns=-1,anchor=south east},
    ]
    \addplot[black, fill=white] table[x=PCC,y=SCC800]{\mydataS};
    \addplot[black, fill=gray] table[x=PCC,y=SCC1800]{\mydataS};
    \addplot[black, fill=black] table[x=PCC,y=SCC2600]{\mydataS};
    \legend{SCC 800, SCC 1800, SCC 2600}
        
    \end{axis}
    
	\node[draw] (label) at (3.4cm,3.6cm) {Intra-Band CA};
	\draw[-stealth] (label) -| +(-2.7cm,-1.0cm);
	\draw[-stealth] (label) -| +(2.6cm,-0.8cm);
	\draw[-stealth] (label) -- +(0cm,-1.0cm);

	\node[draw, fill=white] at (-4mm,-4mm) {DUT-B};

\end{sansmath}
\end{tikzpicture}%
    \endgroup%

	\vspace{-10pt}
	\caption{Additional Power Consumption for Carrier Aggregation at Different Frequencies for \DUTA and \DUTB}
	\label{fig:bandsA}
	\vspace{-15pt}
\end{figure}
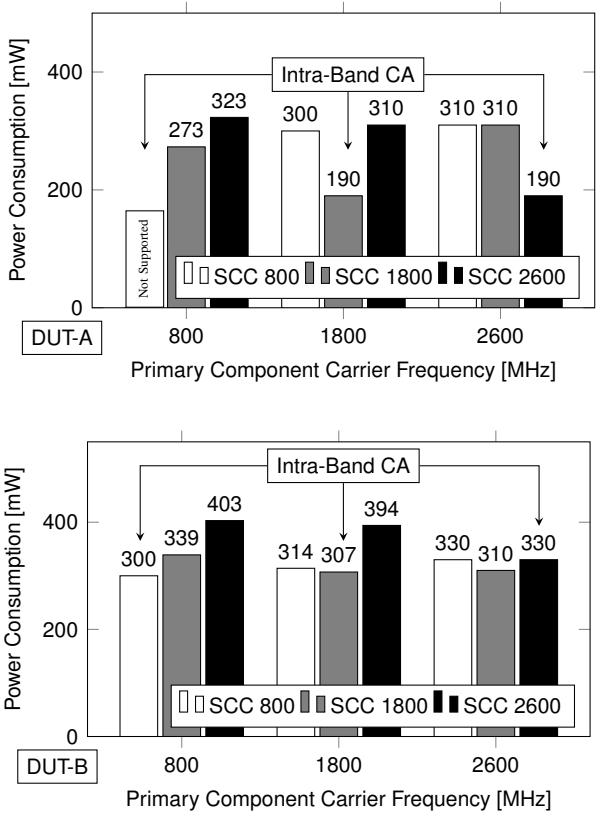

The first measurement targets the question whether it makes a difference in terms of power consumption if a \UE receives the \PCC and \SCC in specific bands and if there are any differences between intra-band and inter-band carrier aggregation.
To quantify this, we sequentially connected the two \DUT to a serving cell in one of three different bands which are licensed for \LTE in Germany:
Band 3 (\SI{1800}{\mega\hertz}),
Band 7 (\SI{2600}{\mega\hertz}),
and Band 20 (\SI{800}{\mega\hertz}).
Subsequently, \SCC at the listed frequencies has been added covering all possible permutations with two carriers and three bands.
The additional power consumption in relation to single carrier operation is plotted in \figurename{~\ref{fig:bandsA}} for both \DUT.

It can be seen, that depending on the selected permutation the additional power consumption of both devices is between \SI{190}{\milli\watt} and \SI{403}{\milli\watt} and varies in the order of \SI{100}{\milli\watt} for a single device.
Particularly, intra-band \CA leads to the smallest power increase with two exceptions:
First, \DUTA does not support any intra-band \CA in the \SI{800}{\mega\hertz} band.
Second, for \DUTB the combination of \SI{2600}{\mega\hertz} and \SI{1800}{\mega\hertz} consumes slightly less power than the corresponding intra-band constellation at \SI{2600}{\mega\hertz}.
For both \DUT with the exclusion of intra-band \CA, enabling an \SCC in the \SI{2600}{\mega\hertz} band results in the highest power draw compared to the other covered bands.

\subsection{Number of Resource Blocks vs. Power Consumption}
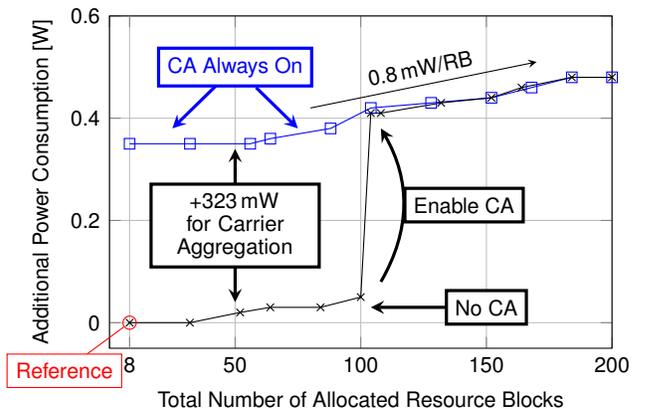
\begin{figure}[tb]  	
	\centering		  
    \begingroup%
    \StrCount{fig/consumption-by-RB-G5}{/}[\matches]%
    \StrBefore[\matches]{fig/consumption-by-RB-G5}{/}[\datapath]%
    \begin{tikzpicture}[spy using outlines={circle, magnification=4, connect spies}]
\begin{sansmath}
\begin{axis}[
		width= 235.0pt,			%
		height=6cm,
		xmin=0, xmax=200,
		ymin=-0.05, ymax=0.6,
		scaled x ticks=false,		%
		extra x ticks={8},
		xlabel={Total Number of Allocated Resource Blocks},
		ylabel={Additional Power Consumption [\si{\watt}]},
		ylabel near ticks,
		grid=both,
		legend cell align=left,
		legend pos=south east,
        ]
        
\addplot[color=blue, mark=square] file {\datapath/consumption-by-RB-G5-PCC+SCC.mat};
\addplot[color=black, mark=x] file {\datapath/consumption-by-RB-G5-PCC.mat};

\tikzstyle{mylabel} = [draw,fill=white]

\draw[stealth-stealth, black, very thick] (axis cs:50,0.04) to node[midway, mylabel, text width=2.0cm, align=center] {+\SI{323}{\milli\watt}\\for Carrier Aggregation} (axis cs:50,0.34);
\draw[-stealth, very thick, bend right] (axis cs:108,0.08) to node[right, mylabel] {Enable CA} (axis cs:108,0.38);

\draw[stealth-, very thick] (axis cs:104,0.03) to node[at end, mylabel] {No CA} (axis cs:150,0.03);
\draw[stealth-, very thick, blue] (axis cs:25,0.38) to node[at end, mylabel, name=oo] {CA Always On} (axis cs:50,0.5);
\draw[-stealth, very thick, blue] (oo) -- (axis cs: 75,0.38);

\node[circle,inner sep=0pt,minimum height=5pt, draw,red] (ref) at (axis cs:8,0) {};

\draw[-stealth] (axis cs:80,0.42) to node[above, sloped] {\SI{0.8}{\milli\watt}/RB} +(axis cs: 90, 0.04) {};

\end{axis}
\draw[red] (ref) to node[at end, draw, fill=white] {Reference} (-.6cm, -.3cm);

\end{sansmath}
\end{tikzpicture}%
    \endgroup%

	\vspace{-10pt}
	\caption{Normalized Power Consumption of \DUTA in Relation to the Number of Allocated Resource Blocks and Carrier Aggregation}
	\label{fig:RBs}
	\vspace{-5pt}
\end{figure}

The second measurement aims at the cause of the power draw increase, whether it is caused by the increased decoding effort or by powering an additional receive chain.
For this purpose we performed two power measurement series for various numbers of allocated \RBs with both, \PCC and \SCC, in the \SI{2600}{\mega\hertz} band at \SI{20}{\mega\hertz} bandwidth.
This allows a total downlink allocation of maximum \num{200} \RBs at the same time.
The first series allocates sequentially \num{8} to \num{100} \RBs on the \PCC only. 
Afterwards \SCC is being added and the sequence is appended to \num{200} \RBs.
The second series covers the same range but enables the \SCC right from the beginning and shares the allocated \RBs equally among the two \CC.
Fig.~\ref{fig:RBs} shows the results of these measurements for \DUTA normalized to the \PCC-only consumption at \num{8}~\RBs.
While the power consumption linearly grows with an approximated rate of \SI{0.8}{\milli\watt} per \RB, adding the \SCC constantly causes a jump of \SI{323}{\milli\watt} independent of the decoding complexity.
As soon as multiple carriers are activated, the distribution of allocated \RBs among the carriers makes no difference in the power consumption.
At this place it must be noted that in contrast to \figurename{~\ref{fig:bandsA}} the operating bandwidth now is twice as high as before, thus the additional power consumption for \CA has roughly doubled as well.
Based on this knowledge, the base station should always prefer -- if possible without the loss of service quality -- a single carrier allocation over an aggregated allocation as long as the number of \RBs fits into the \PCC.
Due to space restrictions the results for \DUTB are omitted as it behaves similarly to \DUTA.

\subsection{Uplink Power Consumption}\label{ch:uplinkpower}
\begin{figure}[tb]  	
	\centering		  
    \begingroup%
    \StrCount{fig/power-vs-tx-power-meas}{/}[\matches]%
    \StrBefore[\matches]{fig/power-vs-tx-power-meas}{/}[\datapath]%
    \begin{tikzpicture}[spy using outlines={circle, magnification=4, connect spies}]
\begin{sansmath}

\begin{axis}[
	width= 235.0pt,			%
	height=4cm,
	xmin=-11, xmax=24,
	ymin=0.4, ymax=2,
	scaled x ticks=false,		%
	xlabel={Transmission Power [dBm]},
	ylabel={Power Cons. [\si{\watt}]},
	ylabel near ticks,
	grid=both,
	legend cell align=left,
	legend pos=north west,
	]

	\tikzstyle{label} = [draw, fill=white]

	\addplot[color=black, mark=square] file {\datapath/power-vs-tx-power-meas-S5.mat};
	\addlegendentry{Measured}
	\addplot[color=blue, dotted, thick] file {\datapath/power-vs-tx-power-approx-high-S5.mat};
	\addlegendentry{Linear Model}
	\addplot[color=blue, dotted, thick] file {\datapath/power-vs-tx-power-approx-low-S5.mat};

	\draw[-] (axis cs: 12, 0) to node[label, near end] {$\gamma$} (axis cs:12, 2);
	\draw[stealth-] (axis cs: 0, 0.606) to node[label, at end] {$\bar{P}_2$} +(0cm,0.5cm);
	\draw[stealth-] (axis cs: 17.2, 1.309) to node[label, at end] {$\bar{P}_3$} +(0cm,-0.5cm);
	\draw[stealth-] (axis cs: 23, 1.873) to node[label, at end, yshift=-1mm] {$\bar{P}_4$} +(-0.5cm,0cm);

\end{axis}

\end{sansmath}
\end{tikzpicture}%
    \endgroup%

	\vspace{-10pt}
	\caption{Power Consumption of \DUTB at \SI{800}{\mega\hertz} in Relation to Transmission Power.}
	\label{fig:txpower}
	\vspace{-5pt}
\end{figure}
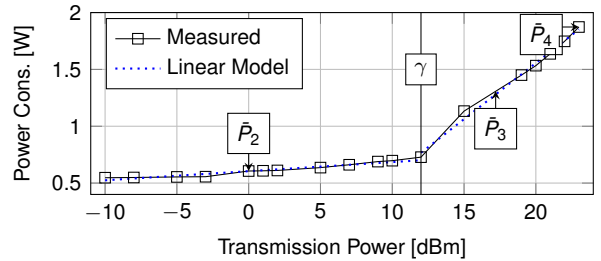

As mentioned in Sec.~\ref{ch:model} the highest power draw of \LTE equipment is caused by the uplink.
Although the focus of this paper lies on the analysis of the downlink, realistic usage patterns of \LTE devices always contain a certain fraction of uplink traffic.
Consequently, a simulation of the over-all power consumption of a \UE requires power measurements in the distinct power states (cf. \figurename{~\ref{fig:statemachine}}).
Since the uplink consumption has been examined in previous works~\cite{Dusza2013b}, only a brief overview of our results is given. %
Measurements were made for both \DUT at \SI{800}{\mega\hertz} and \SI{2600}{\mega\hertz} with a bandwidth of \SI{10}{\mega\hertz} at full \RB allocation.
The \UE transmission power was set by the \BSE covering the range of \SI{-10}{\dBm} to \SI{23}{\dBm} while capturing the power draw of the \DUT.
As an example the measurement results for \DUTB at \SI{800}{\mega\hertz} are given in \figurename{~\ref{fig:txpower}}.
For the power model the consumption is approximated by two linear graphs.
These were separated by the breakpoint $\gamma$ which marks the switching point from a low power amplifier to a high power amplifier within the device~\cite{PhdDusza2013}.
While the power consumptions of the \Idle state $\bar{P}_1$ and the \Max state  $\bar{P}_4$ are clear, the representative consumption for \Low mode $\bar{P}_2$ is taken at \SI{0}{\dBm} and for $\bar{P}_3$ (\High) at $(\gamma+23)/2$ \si{\dBm}~\cite{PhdDusza2013, Dusza2013b}.
The determined values for both \DUT are listed in Tab.~\ref{tab:powers}.

\begin{table}[!t]
	\renewcommand{\arraystretch}{1.3}
	\caption{Empirical Model Parameters for Power Amplifier Modes}
	\label{tab:powers}
	\centering
	\begin{tabular}{ccccc}
		\hline
		Parameter & \multicolumn{2}{c}{\DUTA} & \multicolumn{2}{c}{\DUTB}\\
		\hline
		Frequency [\si{\mega\hertz}] & 800 & 2600 & 800 & 2600\\
		\hline
		$\bar{P}_1$ [\si{\milli\watt}] & 97 & 97 & 30 & 30 \\
		$\bar{P}_2$ [\si{\milli\watt}] & 753 & 860 & 604 & 980 \\
		$\bar{P}_3$ [\si{\milli\watt}] & 1912 & 1578 & 1309 & 1515 \\
		$\bar{P}_4$ [\si{\milli\watt}] & 3053 & 2450 & 1873 & 1993 \\
		\hline
		$\gamma$ [\si{\dBm}] & 15 & 10 & 12 & 12 \\
		\hline
	\end{tabular}
	\vspace{-5pt}
\end{table}

\section{Simulations}\label{ch:simulations}
Based on measurements in Sec.~\ref{ch:measurements}, which we fed into the proposed power consumption model (cf. Sec.~\ref{ch:camodel}), we performed simulations to evaluate the power-savings potential of \CA in various scenarios.
If not stated otherwise the simulations are configured as listed in Tab.~\ref{tab:simuconfig}.

\begin{table}[!t]
	\renewcommand{\arraystretch}{1.3}
	\caption{Default Settings for Simulations}
	\label{tab:simuconfig}
	\centering
	\begin{tabular}{ccccc}
		\hline
		Parameter & Value\\
		\hline
		Frequency & \SI{800}{\mega\hertz}, \SI{2600}{\mega\hertz}\\
		Bandwidth \PCC, \SCC & \SI{10}{\mega\hertz} each\\
		\CA Type & Intra-Band-\CA\\
		\CA Boost Factor $a_\text{CA}$ & 2\\
		Antenna Configuration & \SISO\\
		\hline
		Environment & Suburban\\
		Channel Model & \AWGN\\
		\hline
		Arrival Rate $\lambda$ & $1/(\SI{5}{\min})$\\
		File Size $D_\text{DL}$ & \SI{1}{\giga\bit}\\
		Transmission Protocol & TCP\\
		UL/DL Ratio $d_\text{DLUL}$ & 0.02\\
		UL/DL Rate Ratio $r_\text{DLUL}$ & 0.5\\
		\hline
	\end{tabular}
	\vspace{-5pt}
\end{table}

\subsection{Impact of Data Size}
\begin{figure}[tb]  	
	\centering		  
    \begingroup%
    \StrCount{fig/var-DataSize-G5}{/}[\matches]%
    \StrBefore[\matches]{fig/var-DataSize-G5}{/}[\datapath]%
    \begin{tikzpicture}[spy using outlines={circle, magnification=4, connect spies}, font=\sffamily\footnotesize]
\begin{sansmath}
\begin{axis}[
		width= 235.0pt,			%
		height=5.5cm,
		xmin=1e5, xmax=1e10,
		ymin=0, ymax=1.1,
		xmode=log,
		xticklabels={~,1M,10M,100M,1G,10G},
		scaled x ticks=false,		%
		xlabel={Data Size $D_\text{DL}$ [Bit]}, 
		ylabel={Power Consumption [\si{\watt}]},
		ylabel near ticks,
		grid=both,
		legend cell align=left,
		legend pos=north west,
        ]

\addplot[color=black] table[x index=0, y index=5] {\datapath/var-DataSize.mat};
\addlegendentry{Single Carrier}
\addplot[color=black, dashed] table[x index=0, y index=6] {\datapath/var-DataSize.mat};
\addlegendentry{Carrier Aggregation}
\addplot[color=blue] table[x index=0, y index=7] {\datapath/var-DataSize.mat};
\addplot[color=blue, dashed] table[x index=0, y index=8] {\datapath/var-DataSize.mat};

\draw[-] (axis cs: 1e6, 0.25) to node[draw, fill=white, at start] {\SI{2600}{\MHz}} node[draw, circle, at end, minimum height=5mm] {} (axis cs: 4e7, 0.25);
\draw[-] (axis cs: 2e9, 0.15) to node[draw, fill=white, at start] {\SI{800}{\MHz}} node[draw, circle, at end, minimum height=5mm] {} (axis cs: 9e7, 0.15);

\draw[{stealth}-{stealth}] (axis cs: 1e9, 0.381) to (axis cs: 1e9, 0.515);
\draw[{stealth}-{stealth}] (axis cs: 1e9, 0.564) to (axis cs: 1e9, 0.812);

\draw[-] (axis cs: 1e9, 0.45) to node[at end, draw, fill=white] {\SI{-26}{\percent}} +(-3cm, 0cm);
\draw[-] (axis cs: 1e9, 0.65) to node[at end, draw, fill=white] {\SI{-31}{\percent}} +(-3cm, 0cm);

\end{axis}

\node[draw, fill=white] at (-4mm,-4mm) {DUT-A};

\tikzstyle{mylabel} = [draw,fill=white]

\end{sansmath}
\end{tikzpicture}%
    \endgroup%

	\vspace{5pt}

    \begingroup%
    \StrCount{fig/var-DataSize-S5}{/}[\matches]%
    \StrBefore[\matches]{fig/var-DataSize-S5}{/}[\datapath]%
    \begin{tikzpicture}[spy using outlines={circle, magnification=4, connect spies}, font=\sffamily\footnotesize]
\begin{sansmath}
\begin{axis}[
		width= 235.0pt,			%
		height=5.5cm,
		xmin=1e5, xmax=1e10,
		ymin=0, ymax=1.1,
		xmode=log,
		xticklabels={~,1M,10M,100M,1G,10G},
		scaled x ticks=false,		%
		xlabel={Data Size $D_\text{DL}$ [Bit]},
		ylabel={Power Consumption [\si{\watt}]},
		ylabel near ticks,
		grid=both,
		legend cell align=left,
		legend pos=north west,
        ]
        
\addplot[color=black] table[x index=0, y index=1] {\datapath/var-DataSize.mat};
\addlegendentry{Single Carrier}
\addplot[color=black, dashed] table[x index=0, y index=2] {\datapath/var-DataSize.mat};
\addlegendentry{Carrier Aggregation}
\addplot[color=blue] table[x index=0, y index=3] {\datapath/var-DataSize.mat};
\addplot[color=blue, dashed] table[x index=0, y index=4] {\datapath/var-DataSize.mat};

\draw[-] (axis cs: 1e6, 0.15) to node[draw, fill=white, at start] {\SI{2600}{\MHz}} node[draw, circle, at end, minimum height=5mm] {} (axis cs: 3e7, 0.15);
\draw[-] (axis cs: 2e9, 0.15) to node[draw, fill=white, at start] {\SI{800}{\MHz}} node[draw, circle, at end, minimum height=5mm] {} (axis cs: 2e8, 0.15);

\draw[{stealth}-{stealth}] (axis cs: 1e9, 0.304) to (axis cs: 1e9, 0.383);
\draw[{stealth}-{stealth}] (axis cs: 1e9, 0.619) to (axis cs: 1e9, 0.887);

\draw[-] (axis cs: 1e9, 0.34) to node[at end, draw, fill=white] {\SI{-21}{\percent}} +(-3cm, 0cm);
\draw[-] (axis cs: 1e9, 0.68) to node[at end, draw, fill=white] {\SI{-30}{\percent}} +(-3cm, 0cm);

\end{axis}

\node[draw, fill=white] at (-4mm,-4mm) {DUT-B};

\tikzstyle{mylabel} = [draw,fill=white]

\end{sansmath}
\end{tikzpicture}%
    \endgroup%

	\vspace{-10pt}
	\caption{Power Consumption of \DUTA and \DUTB in Relation to Average Transmission File Size (cf.~Tab.~\ref{tab:simuconfig})}
	\label{fig:var-data-size-G5}
	\vspace{-5pt}
\end{figure}
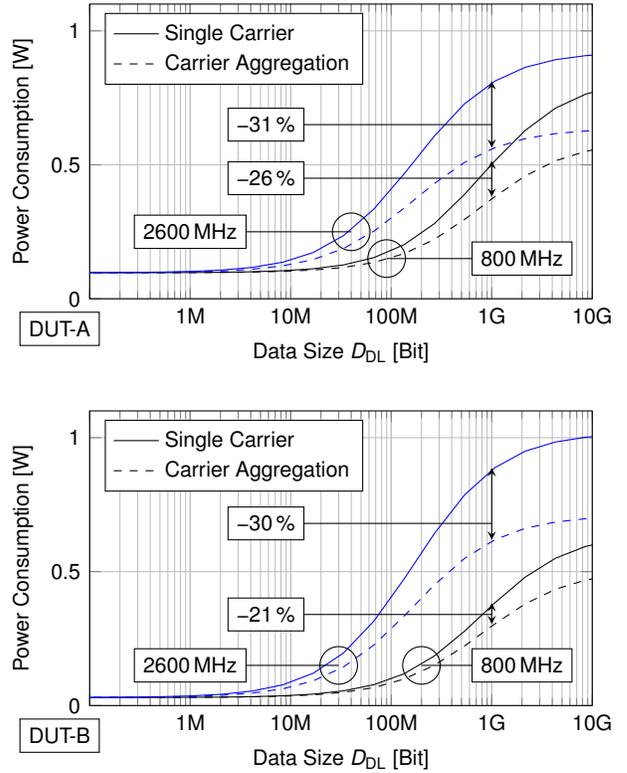

The first simulation varies the average download file size $D_\text{DL}$ for both \DUT at \SI{800}{\mega\hertz} and \SI{2600}{\mega\hertz} as shown in \figurename{~\ref{fig:var-data-size-G5}}.
At small amounts of data ($D_\text{DL} < \SI{2}{\mega\bit}$) most of the power consumption is issued by the idle state, thus being independent of frequency and aggregation type.
For larger transmissions raises the \SI{2600}{\mega\hertz} curve much earlier to a higher power level than the curve of the lower band.
This is mainly caused by the uplink fraction of the transmission and the increased attenuation in this band, which requires in average a higher transmission power of the \UE.
Since both devices consume more power during pure data reception at higher frequencies (cf. $\bar{P}_2$ in Tab.~\ref{tab:powers}), \CA is especially beneficial as the expensive active time is shortened by the boosted data rate and consequently allows the device to fall back quickly into \Idle state.
Both devices consume at this band and at a mean file size of \SI{1}{\giga\bit} about \SI{30}{\percent} less power if \CA is added.
Changing the frequency to \SI{800}{\mega\hertz} results in savings of at least \SI{21}{\percent} to \SI{26}{\percent} depending on the device.

\subsection{Impact of Cell Environment and Boost Factor}
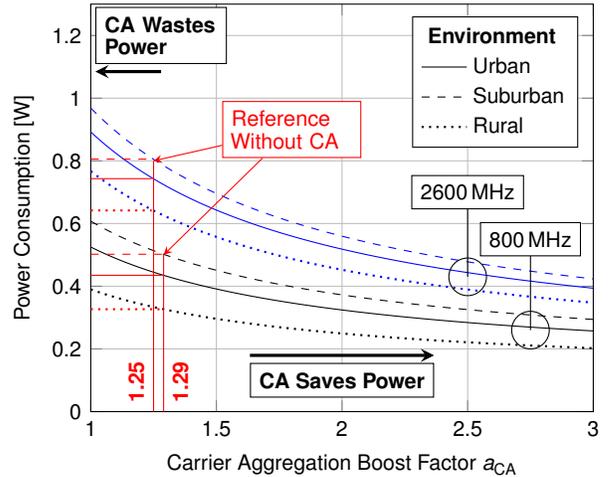
\begin{figure}[tb]  	
	\centering		  
    \begingroup%
    \StrCount{fig/var-CA-boost-G5}{/}[\matches]%
    \StrBefore[\matches]{fig/var-CA-boost-G5}{/}[\datapath]%
    \begin{tikzpicture}[spy using outlines={circle, magnification=4, connect spies}, font=\sffamily\footnotesize]
\begin{sansmath}
\begin{axis}[
		width= 235.0pt,			%
		height=7cm,
		xmin=1, xmax=3,
		ymin=0, ymax=1.3,
		scaled x ticks=false,		%
		xlabel={Carrier Aggregation Boost Factor $a_\text{CA}$},
		ylabel={Power Consumption [\si{\watt}]},
		ylabel near ticks,
		grid=both,
		legend cell align=left,
		legend pos=north east,
        ]

\addlegendimage{empty legend}
\addlegendentry{\hspace{-0.6cm}\textbf{Environment}}   
\addplot[color=black] table[x index=0, y index=1] {\datapath/var-CA-boost-G5.mat};
\addlegendentry{Urban}
\addplot[color=black, dashed] table[x index=0, y index=2] {\datapath/var-CA-boost-G5.mat};
\addlegendentry{Suburban}
\addplot[color=black, dotted, thick] table[x index=0, y index=3] {\datapath/var-CA-boost-G5.mat};
\addlegendentry{Rural}

\addplot[color=blue] table[x index=0, y index=4] {\datapath/var-CA-boost-G5.mat};
\addplot[color=blue, dashed] table[x index=0, y index=5] {\datapath/var-CA-boost-G5.mat};
\addplot[color=blue, dotted, thick] table[x index=0, y index=6] {\datapath/var-CA-boost-G5.mat};

\addplot[color=red, restrict x to domain=0:1.29] table[x index=0, y index=7] {\datapath/var-CA-boost-G5.mat};
\addplot[color=red, dashed, restrict x to domain=0:1.29] table[x index=0, y index=8] {\datapath/var-CA-boost-G5.mat};
\addplot[color=red, dotted, thick, restrict x to domain=0:1.29] table[x index=0, y index=9] {\datapath/var-CA-boost-G5.mat};

\addplot[color=red, restrict x to domain=0:1.25] table[x index=0, y index=10] {\datapath/var-CA-boost-G5.mat};
\addplot[color=red, dashed, restrict x to domain=0:1.25] table[x index=0, y index=11] {\datapath/var-CA-boost-G5.mat};
\addplot[color=red, dotted, thick, restrict x to domain=0:1.25] table[x index=0, y index=12] {\datapath/var-CA-boost-G5.mat};

\draw[-,red] (axis cs:1.25, 0.8) -- (axis cs:1.25, 0);
\draw[-,red] (axis cs:1.29, 0.5) -- (axis cs:1.29, 0);

\node[draw, red, fill=white, text width=1.6cm] (pcc) at (axis cs:1.8, 0.9) {Reference Without CA};
\draw[red, -stealth] (pcc) -- (axis cs: 1.25, 0.8);
\draw[red, -stealth] (pcc) -- (axis cs: 1.29, 0.5);

\node[red, above right, rotate=90] at (axis cs:1.25, 0) {\textbf{1.25}};
\node[red, below right, rotate=90] at (axis cs:1.29, 0) {\textbf{1.29}};

\node[draw, black, fill=white, text width=1.5cm] (waste) at (axis cs:1.28, 1.2) {\textbf{CA Wastes Power}};
\draw[black, stealth-, very thick] ([yshift=-0.1cm]waste.south west) -- ([yshift=-0.1cm]waste.south);

\node[draw, black, fill=white] (save) at (axis cs:2, 0.1) {\textbf{CA Saves Power}};
\draw[black, -stealth, very thick] ([yshift=0.1cm]save.north west) -- ([yshift=0.1cm]save.north east);

\draw[-] (axis cs: 2.5, 0.7) to node[draw, fill=white, at start] {\SI{2600}{\MHz}} node[draw, circle, at end, minimum height=5mm] {} (axis cs: 2.5, 0.43);
\draw[-] (axis cs: 2.75, 0.55) to node[draw, fill=white, at start] {\SI{800}{\MHz}} node[draw, circle, at end, minimum height=5mm] {} (axis cs: 2.75, 0.26);

\end{axis}

\tikzstyle{mylabel} = [draw,fill=white]

\end{sansmath}
\end{tikzpicture}%
    \endgroup%

	\vspace{-10pt}
	\caption{Power Consumption of \DUTA in Relation to Carrier Aggregation Boost Factor for Different Cell Environments}
	\label{fig:var-ca-boost-G5}
	\vspace{-15pt}
\end{figure}

To analyze the impact of the cell environment on the \UE power draw, simulations at urban, suburban, and rural building densities  were performed.
In addition, the \CA boost factor $a_\text{CA}$ was varied to identify the point from which \CA outperforms the single \CC consumption.
The results for \DUTA can be seen in \figurename{~\ref{fig:var-ca-boost-G5}}. %
Results of \DUTB are omitted due space limitations and similar behavior.
As reference, the power consumption of particular scenarios without \CA are drawn in red until they intersect with their corresponding \CA curve.
It can be seen, that the total power consumption of a \UE varies with the chosen environment.
Rural environments issue the lowest power consumption due to the high probability for line-of-sight connections.
The highest consumption takes place in the suburban case as the inter-site distance is still very high (cf. Tab.~\ref{tab:raytracing}) but a large fraction of the covered area has no line-of-sight connection, hence the signals suffers of increased attenuation.
In the urban case the reduced cell size decreases the probability for high power states.

Furthermore, the results show that the intersection of power consumption with \CA and without \CA happens at exactly the same boost factor $a_\text{CA}$ regardless of the chosen environment.
Instead, it depends on the device and operating frequency and lies in a range of \num{1.25} to \num{1.29} for \DUTA.
This means, that \CA reduces the power consumption of an \UE if it increases the downlink data rate at least by an order of \SI{25}{\percent} to \SI{29}{\percent}.
In general, at higher frequencies smaller boost factors are required to reduce the power draw of the device.
While higher boost factors lead to further power savings, the savings do not scale linearly with the boost factor, but rather stagnate beyond values of \num{2} to \num{3}.
In any case, a higher boost factor leads to larger power savings of the \UE.

\subsection{Impact of Mobility and Boost Factor}
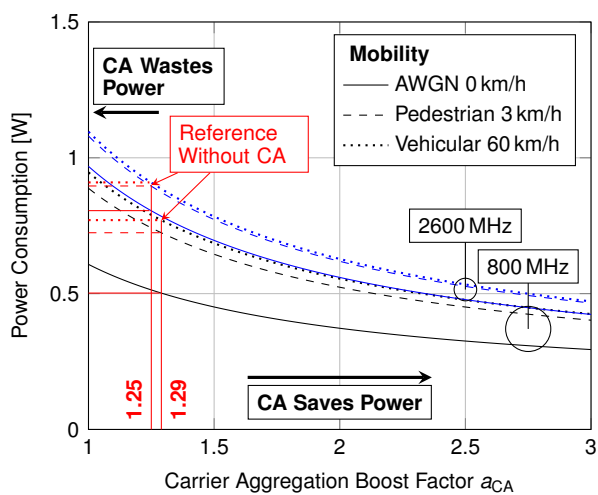
\begin{figure}[tb]  	
	\centering		  
    \begingroup%
    \StrCount{fig/var-CA-boost-channel-G5}{/}[\matches]%
    \StrBefore[\matches]{fig/var-CA-boost-channel-G5}{/}[\datapath]%
    \begin{tikzpicture}[spy using outlines={circle, magnification=4, connect spies}, font=\sffamily\footnotesize]
\begin{sansmath}
\begin{axis}[
		width= 235.0pt,			%
		height=7cm,
		xmin=1, xmax=3,
		ymin=0, ymax=1.5,
		scaled x ticks=false,		%
		xlabel={Carrier Aggregation Boost Factor $a_\text{CA}$},
		ylabel={Power Consumption [\si{\watt}]},
		ylabel near ticks,
		grid=both,
		legend cell align=left,
		legend pos=north east,
        ]

\addlegendimage{empty legend}
\addlegendentry{\hspace{-0.6cm}\textbf{Mobility}}   
\addplot[color=black] table[x index=0, y index=1] {\datapath/var-CA-boost-channel-G5.mat};
\addlegendentry{AWGN \SI{0}{\kilo\meter\per\hour}}
\addplot[color=black, dashed] table[x index=0, y index=2] {\datapath/var-CA-boost-channel-G5.mat};
\addlegendentry{Pedestrian \SI{3}{\kilo\meter\per\hour}}
\addplot[color=black, dotted, thick] table[x index=0, y index=3] {\datapath/var-CA-boost-channel-G5.mat};
\addlegendentry{Vehicular \SI{60}{\kilo\meter\per\hour}}

\addplot[color=blue] table[x index=0, y index=4] {\datapath/var-CA-boost-channel-G5.mat};
\addplot[color=blue, dashed] table[x index=0, y index=5] {\datapath/var-CA-boost-channel-G5.mat};
\addplot[color=blue, dotted, thick] table[x index=0, y index=6] {\datapath/var-CA-boost-channel-G5.mat};

\addplot[color=red, restrict x to domain=0:1.29] table[x index=0, y index=7] {\datapath/var-CA-boost-channel-G5.mat};
\addplot[color=red, dashed, restrict x to domain=0:1.29] table[x index=0, y index=8] {\datapath/var-CA-boost-channel-G5.mat};
\addplot[color=red, dotted, thick, restrict x to domain=0:1.29] table[x index=0, y index=9] {\datapath/var-CA-boost-channel-G5.mat};

\addplot[color=red, restrict x to domain=0:1.25] table[x index=0, y index=10] {\datapath/var-CA-boost-channel-G5.mat};
\addplot[color=red, dashed, restrict x to domain=0:1.25] table[x index=0, y index=11] {\datapath/var-CA-boost-channel-G5.mat};
\addplot[color=red, dotted, thick, restrict x to domain=0:1.25] table[x index=0, y index=12] {\datapath/var-CA-boost-channel-G5.mat};

\draw[-,red] (axis cs:1.25, 0.9) -- (axis cs:1.25, 0);
\draw[-,red] (axis cs:1.29, 0.77) -- (axis cs:1.29, 0);

\node[draw, red, fill=white, text width=1.6cm] (pcc) at (axis cs:1.6, 1.05) {Reference Without CA};
\draw[red, -stealth] (pcc) -- (axis cs: 1.25, 0.9);
\draw[red, -stealth] (pcc) -- (axis cs: 1.29, 0.77);

\node[red, above right, rotate=90] at (axis cs:1.25, 0) {\textbf{1.25}};
\node[red, below right, rotate=90] at (axis cs:1.29, 0) {\textbf{1.29}};

\node[draw, black, fill=white, text width=1.5cm] (waste) at (axis cs:1.28, 1.3) {\textbf{CA Wastes Power}};
\draw[black, stealth-, very thick] ([yshift=-0.1cm]waste.south west) -- ([yshift=-0.1cm]waste.south);

\node[draw, black, fill=white] (save) at (axis cs:2, 0.1) {\textbf{CA Saves Power}};
\draw[black, -stealth, very thick] ([yshift=0.1cm]save.north west) -- ([yshift=0.1cm]save.north east);

\draw[-] (axis cs: 2.5, 0.75) to node[draw, fill=white, at start] {\SI{2600}{\MHz}} node[draw, circle, at end, minimum height=3mm, inner sep=0pt] {} (axis cs: 2.5, 0.515);
\draw[-] (axis cs: 2.75, 0.6) to node[draw, fill=white, at start] {\SI{800}{\MHz}} node[draw, circle, at end, minimum height=6mm] {} (axis cs: 2.75, 0.37);

\end{axis}

\tikzstyle{mylabel} = [draw,fill=white]

\end{sansmath}
\end{tikzpicture}%
    \endgroup%

	\vspace{-10pt}
	\caption{Power Consumption of \DUTA in Relation to Carrier Aggregation Boost Factor for Different Mobility Patterns}
	\label{fig:var-ca-boost-channel-G5}
	\vspace{-15pt}
\end{figure}

After analyzing relationship of \CA and the building density on the power demands of \LTE
 equipment,
now the impact of mobility will be investigated.
Mobility affects the radio channel in terms of fast fading, which is basically a result of the signal's multipath propagation that leads to constructive or destructive interference of the distinct signal paths at the receiver.
As this effect highly depends on position and frequency, the signal fluctuates more intensively if sender or receiver move faster.
Consequently, this reduces the throughput in both directions and requires the \UE to spend more time in active transmission modes.
\figurename{~\ref{fig:var-ca-boost-channel-G5}} shows
the simulated average power consumption of \DUTA for two frequency bands, three mobility types and \CA boost factors from \num{1} to \num{3}.
Again, \DUTB behaves similarly, hence the results are omitted due to limited space.
It can be seen, that mobility -- especially in the \SI{800}{\mega\hertz} band -- significantly increases the power demands of \DUTA if compared to the static \AWGN case in average by \SI{41}{\percent}.
Conversely, increasing the speed of movements from \SI{3}{\kilo\meter\per\hour} to \SI{60}{\kilo\meter\per\hour} increases the power draw in average only by \SI{6}{\percent} at \SI{800}{\mega\hertz}; at \SI{2600}{\mega\hertz} the power increase is even negligible at \SI{1}{\percent}.

The intersection point of power consumption with \CA and without \CA is located at the exactly same boost factor for a given frequency and device independent of the mobility and do not differ from the previous simulation.
This makes \CA beneficial for the battery lifetime of mobile equipment in any case, no matter of the velocity and building density as long as the achieved boost in data rate overbalances the increased \CA power draw.

\section{Conclusion}

The results of this paper show that Carrier Aggregation (CA) is beneficial for the power consumption of \LTE devices when used deliberately.
Measurements of existing \LTEA devices showed that the use of multiple \acf{CC} entails an increased power draw for supplying additional receive chains in the order of \SI{300}{\milli\watt}.
Therefore, as long as the allocated number of \acfp{RB} fits into a single \CC, \CA should be avoided since this only leads to an increased power draw.
This device-specific power increase varies with the chosen band.
In most cases inter-band \CA pointed out a slightly higher power draw than intra-band \CA.

Applying the empirical data into a proposed downlink extension of the \acf{CoPoMo} allowed a quantitative evaluation of the costs and benefits of \CA for different mobility patterns, environments and frequencies.
The simulations show, that \CA has a positive effect on the power consumption, if it boosts the downlink data rate by at least \SI{25}{\percent} to \SI{29}{\percent}.
At this point the power savings by additional idle time compensate the higher power draw during an active transmission.
The exact values depend on the carrier frequency of the primary carrier but not on the building density or whether the \UE is static or moving quickly with a vehicle.

Further simulations showed increasing power savings as the average file size of downloads grows. 
In the simulated scenario at a mean file size of \SI{1}{\giga\bit}, which is typical for multimedia videos or navigation map data, a power reduction of \SI{21}{\percent} to \SI{31}{\percent} can be achieved by \CA.

\section*{Acknowledgment}
Part of the work on this paper has been supported by Deutsche Forschungsgemeinschaft
(DFG) within the Collaborative Research Center SFB 876 ``Providing Information by Resource-Constrained
Analysis'', project A4.

\bibliographystyle{IEEEtran}
\bibliography{Bibliography}
\end{document}